\documentclass[aps,prd,showpacs,floatfix,twocolumn,nofootinbib]{revtex4-1}
\usepackage{amsmath}
\usepackage{amsfonts}
\usepackage{amssymb}
\usepackage{graphicx}
\usepackage[caption=false]{subfig}

\newcommand{\cost}{\cos \left( \theta \right)}
\newcommand{\sint}{\sin \left( \theta \right)}
\providecommand{\adsurl}[1]{\href{#1}{ADS}}

\begin{document}
\title{Visualizing Spacetime Curvature via Gradient Flows III: The Kerr Metric and the Transitional Values of the Spin Parameter}
\author{Majd Abdelqader}
\email[]{majd@astro.queensu.ca}
\author{Kayll Lake}
\email[]{lake@astro.queensu.ca}
\affiliation{Department of Physics, Queen's University, Kingston,
Ontario, Canada, K7L 3N6}
\date{\today}

\begin{abstract}
The Kerr metric is one of the most important solutions to Einstein's field equations, describing the gravitational field outside a rotating black hole. We thoroughly analyze the curvature scalar invariants to study the Kerr spacetime by examining and visualizing their covariant gradient fields. We discover that the part of the Kerr geometry outside the black hole horizon changes qualitatively depending on the spin parameter, a fact previously unknown. The number of observable critical points of the curvature invariants' gradient fields along the axis of rotation changes at several transitional values of the spin parameter. These transitional values are a fundamental property of the Kerr metric. They are physically important since in general relativity these curvature invariants represent the cumulative tidal and frame-dragging effects of rotating black holes in an observer-independent way.

\end{abstract}


\maketitle

\section{Introduction}

The discovery of the Kerr metric in 1963 was a milestone in mathematical relativity and astrophysics. It is a vacuum solution to Einstein's field equations that describes the gravitational field outside a massive axisymmetric rotating object, and no interior solution has been found so far. This solution established that general relativity predicts the existence of the objects we now call black holes, even with angular momentum and away from perfect spherical symmetry. The Kerr metric provides the foundation and framework to study and understand many aspects of modern astrophysics, ranging from supermassive black holes at the center of most galaxies, to supernova explosions and gamma ray bursts. For a thorough review of the Kerr metric and its impact and applications to modern astrophysics, see Ref. \citep{kerrbook}. In Boyer-Lindquist coordinates and using natural units ($G=c=1$), the Kerr metric can be expressed as
\begin{align}\label{kerrmetric}
ds^2=&-\left[ 1-\frac{2mr}{r^2+a^2 \cos^2\theta} \right]dt^2- \frac{4mr\,a\,\sin^2 \theta }{r^2+a^2 \cos^2\theta}\, dt\, d\phi \nonumber \\
 &+\left[\frac{r^2+a^2 \cos^2\theta}{r^2-2mr+a^2} \right]dr^2+(r^2+a^2\cos^2\theta) \, d\theta^2 \nonumber \\
 &+\left[ r^2+a^2+\frac{2mr\,a^2\,\sin^2 \theta }{r^2+a^2 \cos^2\theta} \right]\sin^2\theta \; d\phi^2 ,
\end{align}
where $m$ is the mass, and $a=J/m$, is the angular moment per unit mass, or spin parameter.
\begin{figure}[ht]
\centering
\includegraphics[trim=1.8in 2.5in 1.3in 2.3in, clip, width=2.95in,angle=0]{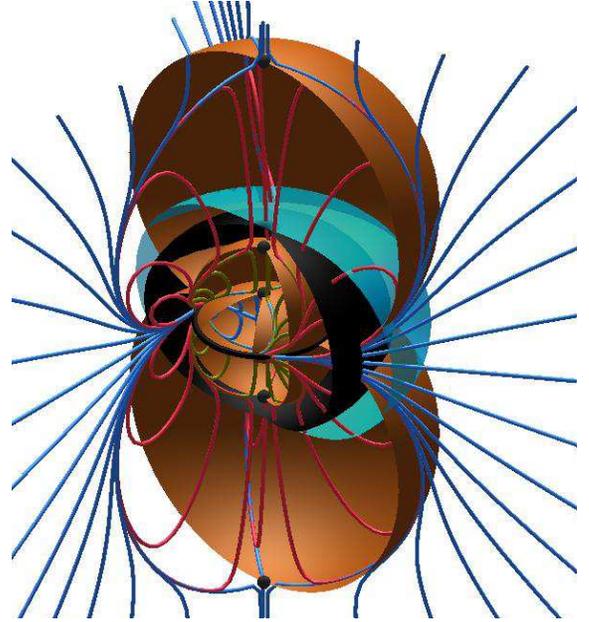}
\caption{(color online). 3-D visualization of rotating black holes in Kerr coordinates via flowlines of the vector field $k_{\mu} \equiv -\nabla_{\mu} (w1R) \;$, for a dimensionless spin parameter value of $A=0.9$. The transparent sky-blue surface is the ergosphere, and the horizon is the black surface. The black circles are the critical points of the gradient field. The colors of the flowlines are of no physical significance, used to distinguish the different regions of the spacetime with common flow structure separated by asymptotic critical directions of the field represented by the brown surfaces, which connect the critical points to the ring singularity.}
\label{w1r3d}
\end{figure}

There is no established standard approach to analyze a metric or a spacetime in general relativity. A common tool used to achieve this is by studying the null and time-like geodesics (i.e. the paths photons and free falling test particles follow) produced by the metric. This approach has been applied extensively to the Kerr metric (see Ref. \citep{kerrbook} and references within). In this paper we explore the geometry of the Kerr metric using a novel analysis and visualization tool proposed in \citep{lake1,abdelqader}. Instead of geodesics, we analyze the geodesic deviations (i.e. tidal and frame-dragging effects) produced by the metric through the Weyl curvature tensor. A somewhat similar approach has been proposed \citep{caltech1, caltech2, caltech3, caltech4}, where the electric and magnetic components of the Weyl tensor, representing tidal and frame-dragging effects respectively, are visualized via tendex and vortex flowlines. Those flowlines are produced by an observer-projection of the electric and magnetic tensors into 3-D spacial coordinates, then obtaining the eigenvectors of those projections. Our approach has a common starting point, but in contrast, we base the analysis on observer-independent curvature invariants constructed from the Weyl tensor. The results obtained this way are genuine properties of the underlying geometry, and independent of any observer or coordinate choice. We discover that the cumulative tidal and frame-dragging effects produced by a rotating black hole change qualitatively depending on the dimensionless ratio between its spin parameter and mass, $A\equiv a/m$. There are 7 unique values of $A$ where critical points of the Weyl curvature invariants emerge from the rotating black hole, and become accessible to observers outside its horizon. This is a major difference compared to non-rotating black holes modeled by the Schwarzschild metric (i.e. $A=0$), where there is only one critical point on the bifurcation 2-sphere, so it is never accessible to an observer outside the horizon, nor can it be produced via gravitational collapse \citep{lake1}.

\section{Gradient Fields of the Weyl Invariants}

\begin{figure*}[ht]
\centering
\subfloat[$w1R$]{
\includegraphics[width=3.42in]{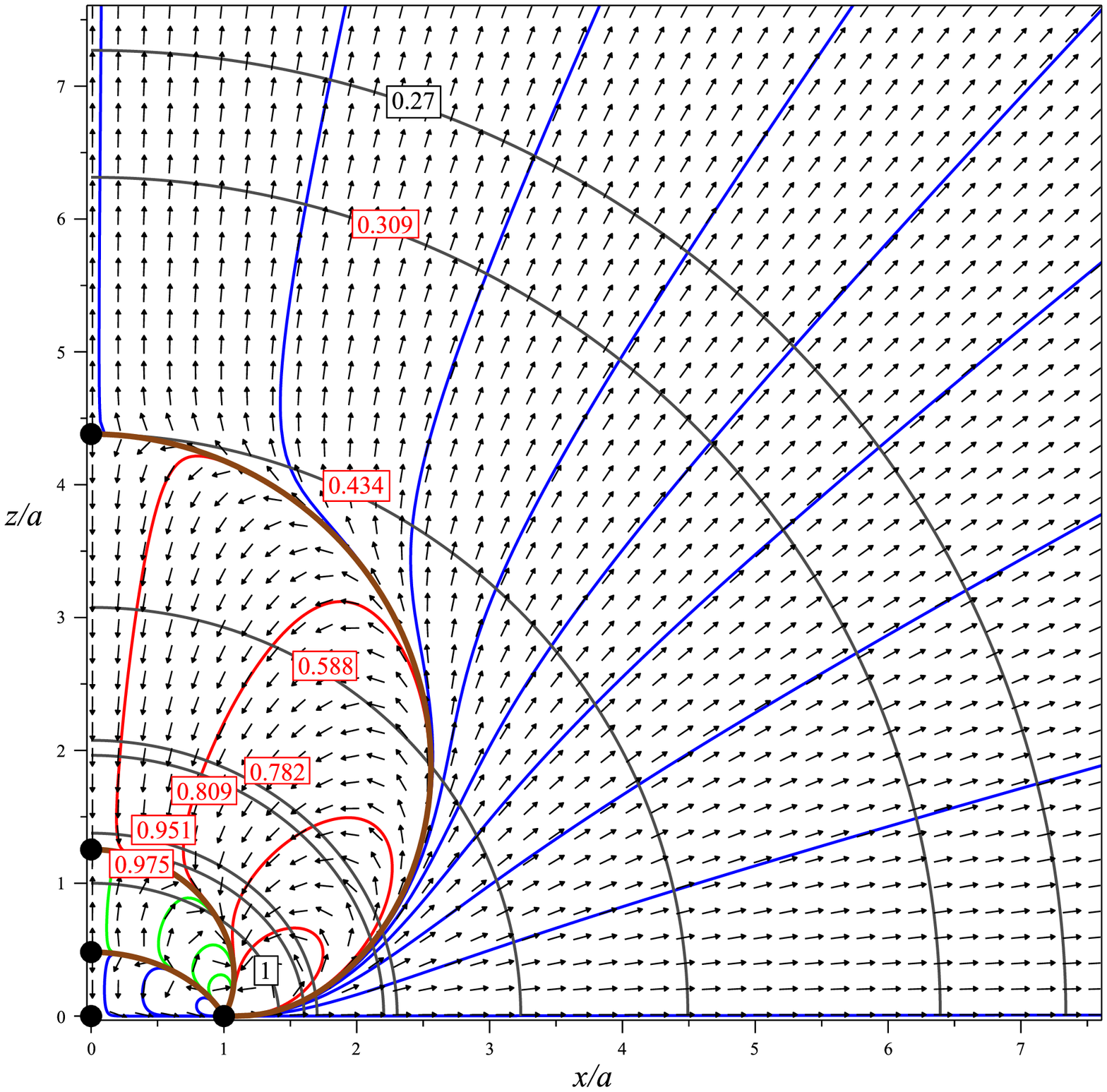}}
\subfloat[$w2R$]{
\includegraphics[width=3.42in]{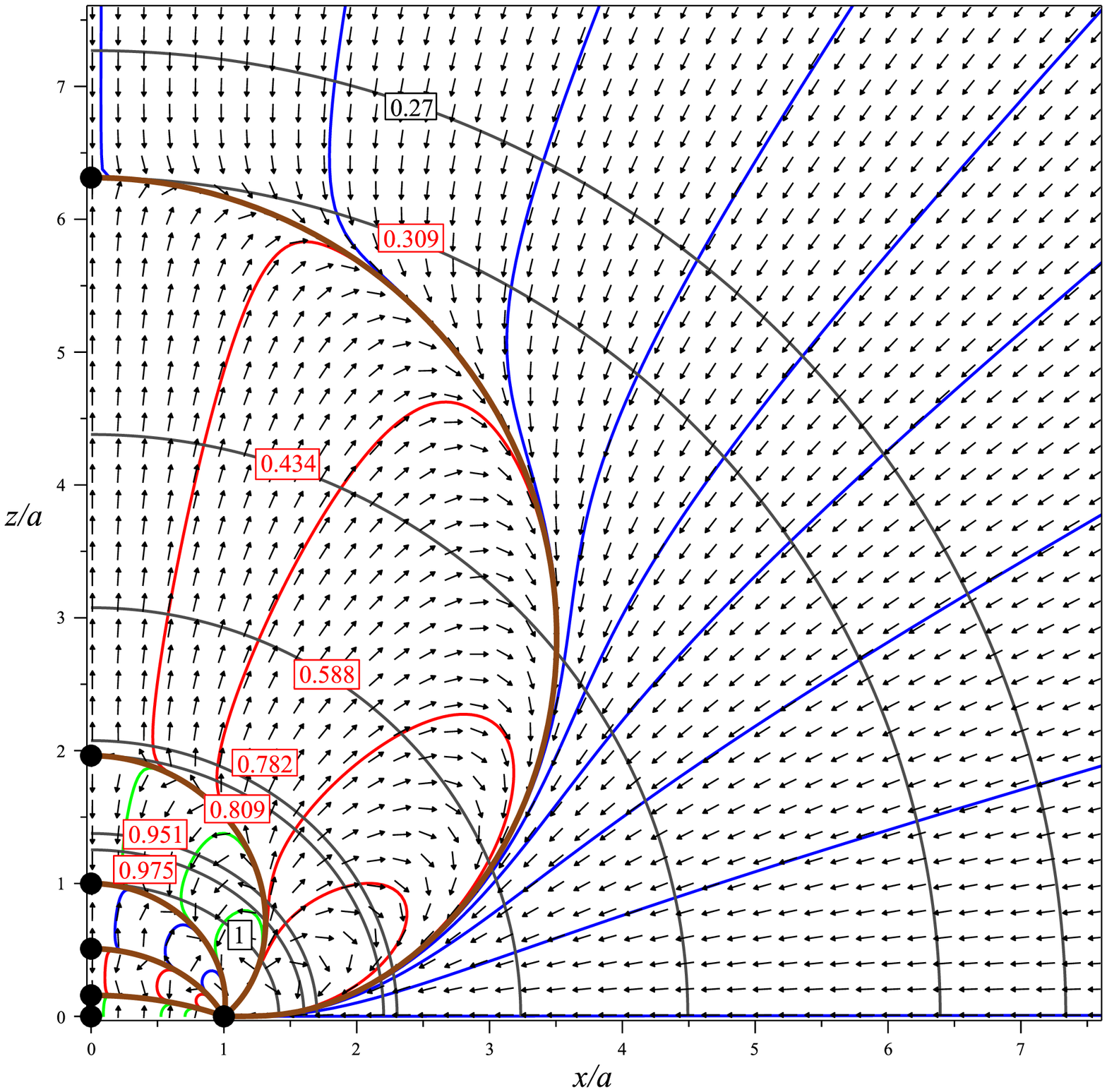}} \\
\vspace{-10pt}
\subfloat[$w1I$]{
\includegraphics[width=3.42in]{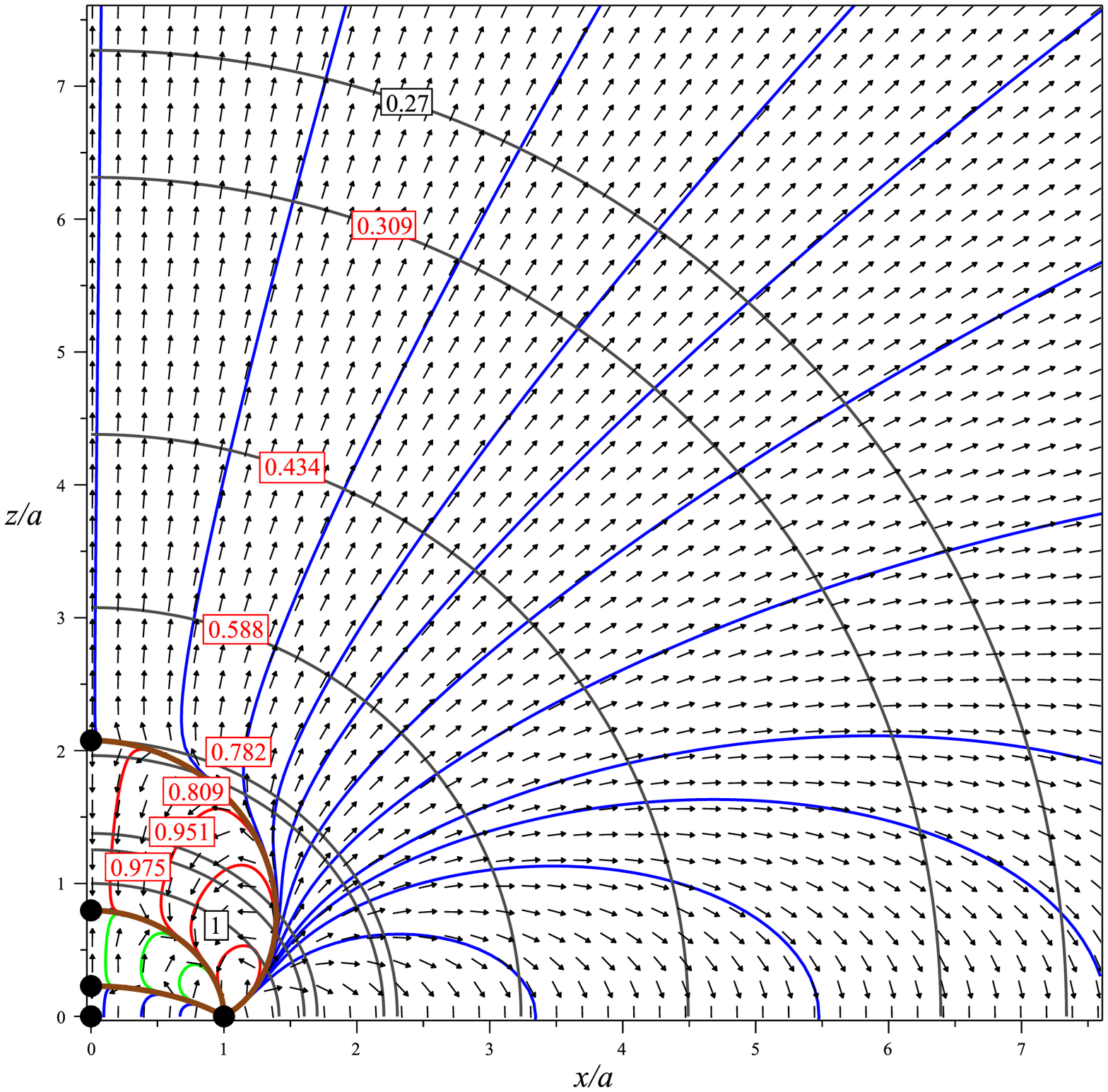}}
\subfloat[$w2I$]{
\includegraphics[width=3.42in]{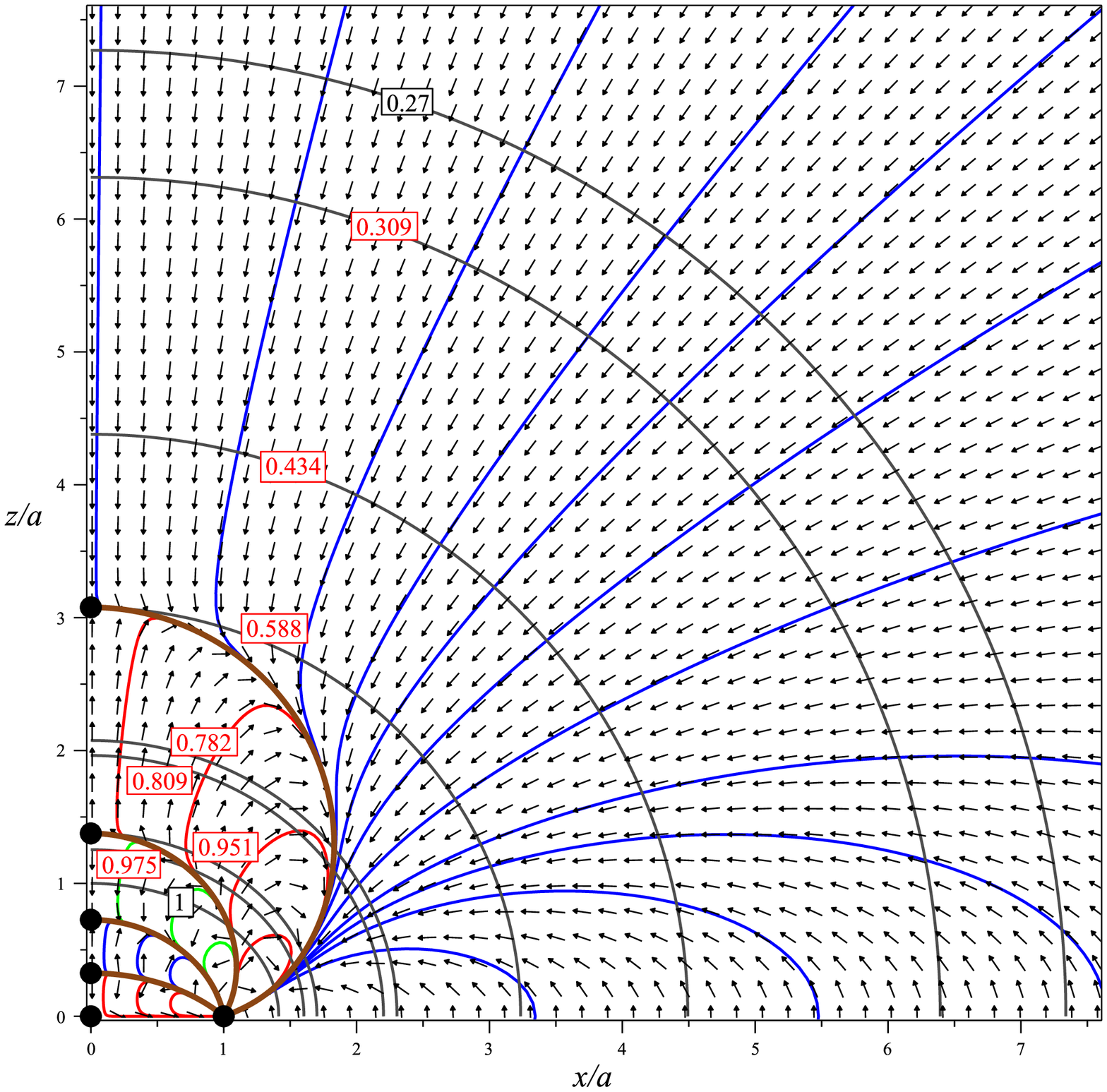}}
\caption{(color online). (a) 2-D slice of Fig. 1 in Kerr-like coordinates introduced in Eq. (\ref{xyz}). The ergosphere is omitted, and different horizons are plotted in grey, for the corresponding transitional values of $A$ indicated in the small boxes, in addition to $A=0.27$ which is not a transitional value but is plotted for presentation purposes. (b), (c), and (d) are the same as (a) but for the gradient vector fields of $w2R$, $w1I$, and $w2I$ respectively.}
\label{fields}
\end{figure*}

As explained in the introductory paper by one of us \citep{lake1}, we limit our analysis to the eight non-differential curvature scalar invariants that have the crucial property of being observer-independent; the four Ricci invariants, and the four Weyl invariants. The Kerr spacetime is Ricci flat, and the four Ricci invariants vanish. The four Weyl invariants for any Petrov type D spacetime, which is the case for the Kerr metric, are algebraically constrained by the complex syzygy $(w1)^3=6(w2)^2\;$, where $w1$ and $w2$ are the complex Weyl invariants, and the four Weyl invariants can be explicitly expressed as $w1R=\Re{(w1)}$, $w2R=\Re{(w2)}$, $w1I=\Im{(w1)}$, and $w2I=\Im{(w2)}$. Note that in vacuum solutions such as the Kerr metric, $w1R=\frac{1}{8} R_{abcd}R^{abcd}=\frac{1}{8}K$, where $K$ is the Kretschmann scalar. However, in order to thoroughly explore the Kerr geometry, it is not sufficient to examine this scalar alone. Furthermore, the syzygy above means that only two of the four invariants are algebraically independent. In spite of this, we find that it is still necessary to analyze the gradient fields for all of the four invariants, because, surprisingly, the structure revealed by each one of these fields is not constrained by the syzygy.

The simplified formulas in Boyer-Lindquist coordinates for each of the invariants are\footnote{This compact way of writing the invariants was pointed out to us by Jan \r{A}man [private communication], upto integer factors to make the invariants consistent with the notation in \cite{lake1}.}

\begin{align}
w1=\frac{6\,{m}^{2}/a^6}{\left( R-i \cos \theta \right) ^{6}} \; , \; \;
 w2=\frac{-6\,{m}^{3}/a^9}{\left( R-i \cos \theta  \right) ^{9}}
\label{invars}
\end{align}
where $R \equiv r/a$. The gradient field of an invariant is simply defined as the covariant derivative $k_{\mu} \equiv -\nabla_{\mu} I = -\partial I/\partial x^{\mu}$, where $I$ is the scalar invariant (see Ref. \cite{lake1}).  The figures presented here are in Kerr coordinates, or pseudo-oblate spheroidal coordinates, inspired by the original Kerr coordinates \footnote{Although the shape of the plotted flowlines depends on the coordinate choice, the existence of each critical point, the classification of each critical point, and the existence of asymptotic critical directions of the fields between critical points is coordinate-independent (See \cite{lake1, abdelqader}).}, where the singularity is unfolded to a ring of radius $a$. Fig. \ref{w1r3d} shows a 3-D visualization of the gradient flow of $w1R$ for the Kerr metric in the original Kerr coordinates. Fig. \ref{fields} shows the gradient fields of the four invariants, as well as the outer horizon for different values of the dimensionless spin parameter $A$. The transformation equations from Boyer-Lindquist coordinates to the coordinates used in Fig. \ref{fields} are
\begin{align}\label{xyz}
x &= \sqrt{r^2+a^2}\, \sint \cos \left( \phi \right) \nonumber \\ 
y &= \sqrt{r^2+a^2}\, \sint \sin \left( \phi \right) \\ \nonumber
z &= r\,\cost \, 
\end{align}
where the only difference between these coordinates and the original Kerr coordinates is that $[\phi]$ is replaced by $\left[ \phi -\tan^{-1} \left( a/r \right) \right]$ in the equations above for $x$ and $y$. We choose to omit the $\left[ \tan^{-1}(a/r) \right]$ term in Fig. \ref{fields} to keep the gradient fields $\phi$-independent and allow us to plot a 2-D slice of the spacetime with constant $\phi$. The spacetime in these coordinates is symmetric about the $z$-axis, and about the $x-y$ plane, therefore it is sufficient to plot one quadrant of the 2-D slice of the spacetime with constant $\phi$ and $t$, and in this case we use $\phi=0$ (i.e. the $z-x$ plane).

With the exception of the singularity, all of the critical points of the four gradient fields (i.e. the points where $k_{\mu}=0$) lie on the z-axis \footnote{For a thorough definition and classification method of these critical points in general, see Ref. \cite{lake1}.}. There are exactly 7 critical points with $z>a$, more specifically $z_{critical}/a$=[6.314, 4.381, 3.078, 2.077, 1.963, 1.376, 1.254] and another 7 critical points with $z<a$, where $z_{critical}/a$=[0.7975, 0.7265, 0.5095, 0.4816, 0.3249, 0.2282, 0.1584]. The 14 values of $z_{critical}$ above are evaluated numerically to 4 significant figures.

\section{Transitional Values of the Spin Parameter}

\begin{figure}[th]
\centering
\includegraphics[width=3.4in]{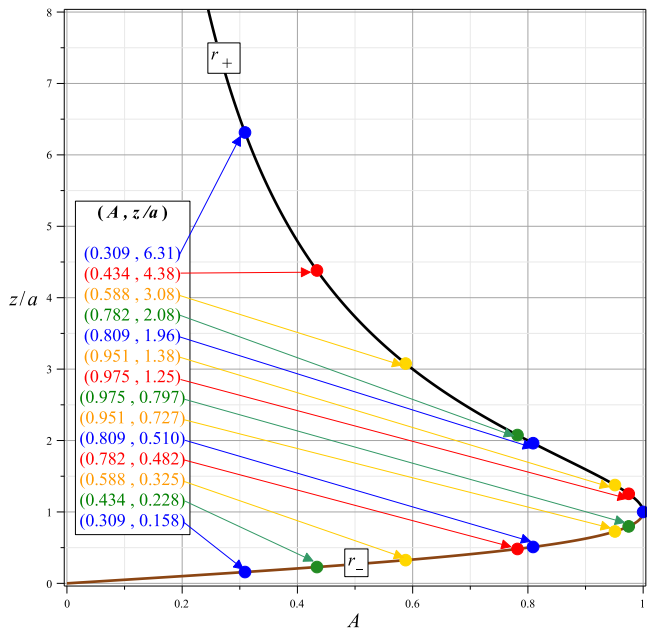}
\caption{(color online). The outer horizon $r_{+}$ (black) and the inner horizon $r_{-}$ (brown) of a rotating black hole vs. its dimensionless spin parameter $A$. The transitional values of $A$ are indicated by the red, blue, green, and yellow points, associated with the horizons crossing one of the critical points of the invariant $w1R$, $w2R$, $w1I$ and $w2I$ respectively.}
\label{horizons}
\end{figure}

The event horizons of the Kerr black hole are given by $r_{\pm}=m \pm \sqrt{m^2-a^2}$. Therefore, when expressed in the dimensionless form, they simplify to $r_{\pm}/a=R_{\pm}=\left( 1 \pm \sqrt{1-A^2}\, \right)/A$. Note that the horizons still depend on the mass and spin parameter through $A$ in a non-linear way. On the other hand, the locations of the critical points of the gradient fields along the axis of rotation scale linearly with $a$, as we stated above. Therefore, we find that the observable part of the gradient fields of the Weyl scalar invariants of the Kerr metric (i.e. the structure of curvature outside a rotating black hole) changes qualitatively depending the dimensionless spin parameter $A$. Furthermore, at very specific values of $A$, the outer horizon crosses one of the critical points of the four gradient fields, making that critical point accessible to an observer outside the black hole. The transitional values of the spin parameter are $A$=[0.3090, 0.4339, 0.5878, 0.7818, 0.8090, 0.9511, 0.9749], evaluated to 4 significant figures. We also find that the inner horizon also crosses one of the critical points with $z_{critical}<a$ at each one of the transitional values of $A$ stated above. This indicates a deeper connection between the four Weyl invariants and the horizons of the Kerr black hole, even though the algebraic form of the invariants does not seem to include any information about the horizons in Eq. \ref{invars}. Fig. \ref{horizons} summarizes the result, plotting the the outer and inner horizons as a function of $A$. The transitional values of $A$ mentioned above are indicated on the curves, along with the value of $z_{critical}$ that the horizons cross at each transitional $A$.
\section{Conclusion}
We have applied a new curvature visualization and analysis tool to the Kerr metric, and have discovered some fundamental properties of the Kerr spacetime. The observable structure of the Kerr geometry outside the horizon of rotating black holes changes qualitatively at 7 specific values of the spin parameter. The invariants under scrutiny represent the cumulative tidal and frame dragging effects that the rotating black hole exerts on the surroundings, as explained in detail in \citep{lake1,abdelqader,caltech1,caltech2,caltech3,caltech4}. In other words, if one can send a spaceship with very precise tools to measure tidal and frame dragging forces, it is possible in principle to map the values of these invariants around a rotating black hole.  Therefore, the critical points that emerge from the horizon at each transitional value of the spin parameter are observable in principle. It is interesting to note that the region surrounding the event horizon of Sagittarius A* will soon be visible by way of the Event Horizon Telescope \cite{EHT1,EHT2}.

Another useful application of the technique in this paper would be in the analysis of numerical relativity simulations. The visualization tool presented here provides a coordinate-independent, observer-independent, and physically intuitive picture where the flowlines seek the extremum points of curvature associated with each invariant. Furthermore, proper lengths between the critical points (e.g. the null affine distances or the proper length of a time-like curve between points) would provide a coordinate independent way to extract angular moment and mass estimates of black holes in the simulations.

\begin{acknowledgments}
The authors would like to thank Robert Owen for helpful discussions. This work was supported in part by a Duncan and Urlla Carmichael Graduate Fellowship (to MA) and a grant (to KL) from the Natural Sciences and Engineering Research Council of Canada. Portions of this work were made possible by use of \textit{GRTensorII} \cite{grt}.
\end{acknowledgments}

\end{document}